\documentclass[aps,prl,twocolumn,showpacs,byrevtex]{revtex4-2}

\usepackage{times,mathptm}
\usepackage[dvips]{graphicx,color}
\usepackage{titlesec}
\newcommand{\bqa}{\begin{eqnarray*}}
\newcommand{\eqa}{\end{eqnarray*}}

\begin{document}

\title{Hole-Doping Effect on Superconductivity in Compressed CeH$_{9}$ at High Pressure}
\author{Chongze Wang$^1$, Shuyuan Liu$^1$, Hyunsoo Jeon$^1$, Seho Yi$^1$, Yunkyu Bang$^{2,3}$, and Jun-Hyung Cho$^{1,3*}$}
\affiliation{$^1$ Department of Physics, Research Institute for Natural Science, and Institute for High Pressure at Hanyang
University, Hanyang University, 222 Wangsimni-ro, Seongdong-Ku, Seoul 04763, Republic of Korea \\
$^2$ Department of Physics, Pohang University of Science and Technology, Pohang 37673, Republic of Korea \\
$^3$ Asia Pacific Center for Theoretical Physics (APCTP),  Pohang-si, Gyeongsangbuk-do 37673, Republic of Korea}
\date{\today}

\begin{abstract}
The experimental realization of high-temperature superconductivity in compressed hydrides H$_3$S and LaH$_{10}$ at high pressures over 150 GPa has aroused great interest in reducing the stabilization pressure of superconducting hydrides. For cerium hydride CeH$_9$ recently synthesized at 80$-$100 GPa, our first-principles calculations reveal that the strongly hybridized electronic states of Ce 4$f$ and H 1$s$ orbitals produce the topologically nontrivial Dirac nodal lines around the Fermi energy $E_F$, which are protected by crystalline symmetries. By hole doping, $E_F$ shifts down toward the topology-driven van Hove singularity to significantly increase the density of states, which in turn raises a superconducting transition temperature $T_c$ from 74 K up to 136 K at 100 GPa. The hole-doping concentration can be controlled by the incorporation of Ce$^{3+}$ ions with varying their percentages, which can be well electronically miscible with Ce atoms in the CeH$_9$ matrix because both Ce$^{3+}$ and Ce behave similarly as cations. Therefore, the interplay of symmetry, band topology, and hole doping contributes to enhance $T_c$ in compressed CeH$_9$. This mechanism to enhance $T_c$ can also be applicable to another superconducting rare earth hydride LaH$_{10}$.

\end{abstract}

\maketitle

\vspace{0.5cm}


Doping in condensed matters is a well-established means of manipulating their electronic structures, which may lead to the emergence of various quantum phases with exotic physical properties~\cite{Defect-Haller,Tokura-Nature1989,TBG-Nature2018,Rev-cuprate-2006,Rev-cuprate-2015,Rev-pnictide-2011,Rev-pnictide-2015}. For example, in the unconventional high-temperature superconductors such as cuprates~\cite{cuprate-Muller1986} and pnictides~\cite{pnictides-JACS2006,pnictides-JACS2008}, doping by holes or electrons has been demonstrated not only to induce complex quantum phase transitions including magnetism, pseudogap, charge density wave, superconductivity (SC), and Fermi liquid phases, but also to vary $T_c$ in their superconducting phases~\cite{Rev-cuprate-2006,Rev-cuprate-2015,Rev-pnictide-2011,Rev-pnictide-2015}. Due to the emergence of such many electronic states in unconventional high-$T_c$ superconductors, identifying the mechanism responsible for the doping-induced changes of $T_c$ has been elusive. By contrast, doping effect in conventional Bardeen-Cooper-Schrieffer (BCS)~\cite{BCS} superconductors has been relatively well understood in terms of the influence of electron-phonon coupling (EPC), and therefore various dopants can be employed to tune $T_c$. It is thus very interesting and challenging to investigate the effect of doping on the EPC-driven SC of recently discovered hydrides at high pressures~\cite{new-review}.

During the past six years, compressed hydrides under megabar pressures have attracted much attention because of their unprecedented records of $T_c$. Motivated by the theoretical predictions of SC in a number of hydrides~\cite{LiHx-PANS2009,LiHx-acta cryst.2014,KHx-JPCC2012,CaH6-PANS2012,H3S-Sci.Rep2014,MgH6-RSC-Adv.2015,rare-earth-hydride-PRL2017,rare-earth-hydride-PANS2017,Li2MgH16-PRL2019,HfH10-PRL2020}, experiments have confirmed that sulfur hydride H$_3$S and lanthanum hydride LaH$_{10}$ exhibit $T_{\rm c}$ around 203 K at ${\sim}$155 GPa~\cite{ExpH3S-Nature2015} and 250$-$260 K at ${\sim}$170 GPa~\cite{ExpLaH10-PRL2019, ExpLaH10-Nature2019}, respectively. More recently, carbonaceous sulfur hydride was experimentally realized to reach a room-temperature SC with a $T_{\rm c}$ of 288 K at ${\sim}$267 GPa~\cite{Exp-CSH-Nature2020}. Despite such an achievement of room-temperature SC, it is highly demanding to discover high-$T_c$ superconducting hydrides synthesized at moderate pressures below ${\sim}$100 GPa, which can be normally achievable with the diamond anvil cell~\cite{diamondanvil-Rev2009-Bassett,diamondanvil-Rev2018-K.K.Mao}. Near simultaneously, two experimental groups~\cite{ExpCeH9-Nat.Commun2019T.Cui,ExpCeH9-Nat.Commun2019-J.F.Lin} reported the successful synthesis of cerium hydride CeH$_9$ at 80$-$100 GPa. The subsequent density-functional theory (DFT) calculation of CeH$_9$ revealed that the delocalized nature of Ce 4$f$ electrons is an essential ingredient in the high chemical precompression of clathrate H cage around Ce atom [see Fig. 1(a)]. It is noticeable that, even though the synthesis of CeH$_9$ was made at lower pressures below ${\sim}$100 GPa, its theoretically predicted $T_c$ value was around 75 K~\cite{rare-earth-hydride-PRL2017}, much lower than those of H$_3$S and LaH$_{10}$~\cite{ExpH3S-Nature2015,ExpLaH10-PRL2019,ExpLaH10-Nature2019}. Therefore, the main bottleneck for the research of high-$T_c$ superconducting hydrides has been associated with difficulties both raising $T_c$ and lowering the pressure of stability simultaneously. In order to alleviate this bottleneck in CeH$_9$, we here investigate the effect of hole doping on SC, which leads to a significant increase in $T_c$.

For high-pressure rare earth hydrides with clathrate H-cage structures, the electronic states tend to have a strong hybridization between rare earth-4$f$ and H-1$s$ orbitals near $E_{\rm F}$~\cite{rare-earth-hydride-PANS2017,LaH10-liangliang,chongze1,chongze2,CeH9-hyunsoo,LaH10-Papacon}. This electronic characteristic of rare earth hydrides having high-symmetry structures could be favorable for hosting topological states through band inversions, identified in recent studies of topological materials~\cite{topo-summary,shuyuan-electride}. As compelling examples, topologically nontrivial Dirac-nodal-line (DNL) states are jointly protected by the space inversion symmetry $P$ and time-reversal symmetry $T$ supplemented by additional crystalline symmetry~\cite{NL-PRL2015,NL-PRB2015,NL-NC2016,NL-CPB2016}. However, exploration of the cooperative interplay of crystal symmetry and band topology has so far been overlooked in high-pressure superconducting hydrides. These new ingredients of symmetry and topology together with hole doping will provide a promising playground to enhance $T_c$ in high-pressure superconducting hydrides, as will be demonstrated below.

In this Letter, using first-principles calculations, we discover that CeH$_{9}$ possessing a hexagonal-close-packed (hcp) structure has symmetry-enforced DNL states. It is revealed that the two-dimensional (2D) nodal surface guaranteed by the nonsymmorphic crystal symmetry $S_{2z}$ (equivalent to the combination of twofold rotation symmetry $C_{2z}$ about the $z$ axis and a half translation along the $z$ direction) is converted to one-dimensional (1D) DNLs in the presence of spin-orbit coupling (SOC)~\cite{NS-PRB2016}. Moreover, two DNL states composed of strongly hybridized Ce 4$f$ and H 1$s$ orbitals touch each other, leading to the formation of a van Hove singularity (vHs) around $-$1.6 eV below $E_F$. Consequently, hole doping shifts $E_F$ toward the vHs, which in turn increases EPC and therefore raises $T_c$ from 74 K (without hole doping) up to 136 K at 100 GPa. Considering that Ce atoms in the CeH$_9$ matrix behave as cations, Ce$^{3+}$ ions is expected to be well electronically miscible with Ce atoms and their incorporation percentages can control hole-doping concentrations. Our findings provide a new avenue for using hole doping to enhance $T_c$ in recently synthesized rare earth hydrides CeH$_9$~\cite{ExpCeH9-Nat.Commun2019T.Cui,ExpCeH9-Nat.Commun2019-J.F.Lin} as well as LaH$_{10}$~\cite{ExpLaH10-PRL2019, ExpLaH10-Nature2019}.

\begin{figure}[ht]
\centering{ \includegraphics[width=8.5cm]{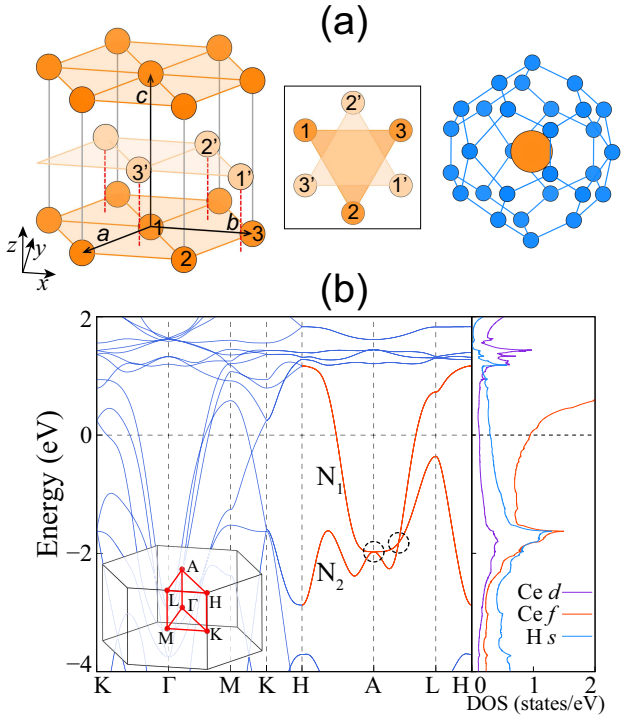} }
\caption{(a) Optimized hcp structure of Ce atoms in CeH$_{9}$. The inset shows the top view of Ce atoms, and the isolated H$_{29}$ cage surrounding a Ce atom is also included. The calculated band structure of CeH$_9$ together with the PDOS for Ce 4$f$, Ce 5$d$, and H 1$s$ orbitals is given in (b). The unit of DOS is states/eV per unit cell that contains two Ce atoms. $N_1$ and $N_2$ represent the fourfold degenerate bands along the high-symmetry $H$-$A$-$L$-$H$ paths, and the energy zero is $E_F$. The Brillouin zone of hcp structure is also included in (b).}
\end{figure}

We first present the electronic band structure of hcp CeH$_9$, obtained using first-principles DFT calculations~\cite{methods}. In most of the calculations hereafter, we fix a pressure of 100 GPa at which the hcp phase with the lattice parameters $a$ = $b$ = 3.698 {\AA} and $c$ = 5.596 {\AA} [see Fig. 1(a)] is thermodynamically stable [see Fig. S1(a) in the Supplemental Material~\cite{SM}]. It is noted that at 70 GPa, the hcp phase becomes dynamically unstable with the presence of imaginary phonon frequencies [see Fig. S1(b)]. Figure 1(b) shows the calculated band structure and partial density of states (PDOS) of CeH$_9$. We find that the Ce 4$f$ and H 1$s$ orbitals are more dominant components in the electronic states around $E_F$, compared to other orbitals (see Fig. S2 in the Supplemental Material~\cite{SM}). Interestingly, the PDOS for Ce 4$f$ and H 1$s$ orbitals exhibits a sharp peak around $-$1.6 eV below $E_F$ [see Fig. 1(b)], indicating a strong hybridization of the two orbitals. The existence of such a vHs having large DOS leads to an increase of $T_c$ via hole doping, as discussed below.

Figure 1(b) represents the DFT band structure computed without including SOC. The presence of $P$ and $T$ symmetries ensures Kramer's double degeneracy in the whole Brillouin zone (BZ). We find that there are fourfold degenerate bands $N_1$ and $N_2$ along the high-symmetry H-A-L-H paths, formed by touching of two bands. It is noted that $N_1$ and $N_2$ touch each other at $A$ and between $A$ and $L$ [marked by dashed circles in Fig. 1(c)], thereby giving rise to eigthfold accidental degeneracies. Using the tight-binding Hamiltonian with a basis of maximally localized Wannier functions~\cite{w90,wt}, we reveal the existence of 2D nodal surfaces $NS_1$ and $NS_2$ throughout the $k_z$ = ${\pi}$/$c$ plane, as shown in Fig. 2(a). Here, each nodal surface is formed by a touching of two doubly-degenerate bands at the boundary of BZ. Since the crystalline symmetry of hcp CeH$_9$ belongs to the space group $P6_3/mmc$ (No. 194) with the point group $D_{6h}$, the fourfold degeneracy of $NS_1$ and $NS_2$ is respected by the combined symmetry $PS_{2z}$, whose eigenvalues are ${\pm}1$ because of ($PS_{2z}$)$^2$ = 1 (see symmetry analysis in the Supplemental Material~\cite{SM}). The inclusion of SOC lifts the degeneracy of $N_1$ and $N_2$ along the $H$-$A$-$L$-$H$ paths except $A$-$L$ (see Fig. S3 in the Suppelemental Material~\cite{SM}), where the SOC-induced gap opening is less than ${\sim}$0.1 eV [see Fig. 2(b)]. It is noted that the nodal surfaces $NS_1$ and $NS_2$ are converted into 1D nodal lines along the high-symmetry paths $k_x$ = 0 and $k_x$ = ${\pm}\sqrt{3}k_y$ as well as with circular patterns around the $A$ point [see Fig. 2(b)]. These DNLs showing $C_{3z}$ rotation symmetry are protected by additional mirror symmetry (see symmetry analysis in the Supplemental Material~\cite{SM}). For example, $M_x$ : ($x$,$y$,$z$) ${\rightarrow}$ ($-x$,$y$,$z$) anticommuting with $PS_{2z}$ allows the existence of the fourfold degenerate nodal line at $k_x$ = 0 on the $k_z$ = ${\pi}$/$c$ plane. The topological characterizations of these DNLs are demonstrated by calculating the topological index~\cite{NL-CPB2016}, defined as ${\zeta}_1$ = ${\frac{1}{\pi}}$ ${\oint}$$_c$ $dk$${\cdot}$A($k$), along a closed loop encircling any of the DNLs. Here, A(k) = $-i$$<$$u_k$$\mid$$\partial$$_k$$\mid$$u_k$$>$ is the Berry connection of the related Bloch bands. We obtain ${\zeta}_1$ = ${\pm}$1 for the DNLs, indicating that they are stable against $PS_{2z}$ and $M$ symmetries conserving perturbations.

\begin{figure}[ht]
\centering{ \includegraphics[width=8.5cm]{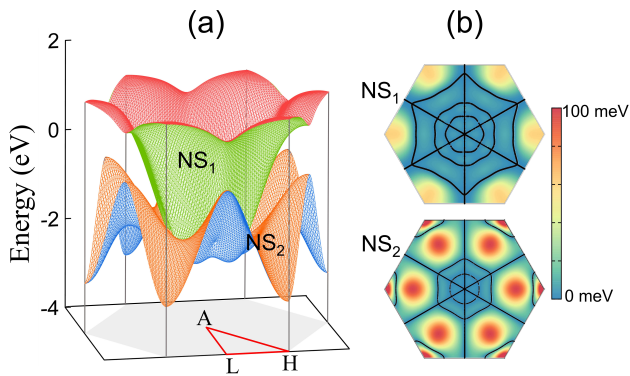} }
\caption{(a) Energy of 2D nodal surfaces $NS_1$ and $NS_2$ throughout the $k_z$ = ${\pi}$/$c$ plane, obtained without including SOC and (b) 1D nodal lines converted from $NS_1$ and $NS_2$ with including SOC. In (b), the SOC-induced gap is represented using the color scale in the range between 0 and 100 meV.}
\end{figure}

As shown in Figs. 1(b) and 2(a), the position of vHs arising from the saddle points of energy dispersion is located near the band touching of $N_1$ ($NS_1$) and $N_2$ ($NS_2$). It is thus likely that crystalline symmetries producing the nontrivial band topology of CeH$_9$ are correlated with the existence of vHs. In other words, the present vHs is rendered emergent by the band touching of the two fourfold degenerate topological states below $E_F$ and yield a power law divergence in the DOS~\cite{vHs}, which in turn enhances $T_c$ via hole doping, as discussed below.

To estimate $T_c$ of CeH$_9$ at 100 GPa, we calculate the phonon spectrum, projected phonon DOS onto Ce and H atoms, Eliashberg function ${\alpha}^{2}F({\omega})$, and integrated EPC constant ${\lambda}({\omega})$ as a function of phonon frequency. Figure 3(a) shows that the phonon spectrum is divided into two regimes: i.e., one is the low-frequency regime arising from the vibrations of Ce atoms and the other is the high-frequency regime driven by H atoms. Therefore, we find that the acoustic phonon modes of Ce atoms contribute to ${\sim}$19\% of the total EPC constant ${\lambda}$ = ${\lambda}$(${\infty}$), whereas the optical phonon modes of H atoms contribute to ${\sim}$81\% of ${\lambda}$. Specifically, the H-derived low-frequency optical modes close to the Ce-derived acoustic modes show larger EPC strength, as represented by circles on the phonon dispersion in Fig. 3(a). Based on these results, we can say that the optical vibrations of H atoms are strongly coupled to the hybridized electronic states of Ce 4$f$ and H 1$s$ orbitals around $E_F$, giving rise to ${\lambda}$ = 1.04. By numerically solving the isotropic Migdal-Eliashberg equations~\cite{Migdal,Eliash,ME-review}, we calculate the superconducting gap versus temperature with varying Coulomb pseudopotential parameter ${\mu}^*$~\cite{rare-earth-hydride-PRL2017,ExpCeH9-Nat.Commun2019-J.F.Lin}, and estimate $T_c$ ${\approx}$ 84 and 74 K with ${\mu}^*$ = 0.1 and 0.13, respectively [see Fig. 3(b)]~\cite{ref-SOC}. These predicted $T_c$ values of CeH$_9$ are much lower than the experimentally observed $T_c$ ${\approx}$ 260 K of LaH$_{10}$~\cite{ExpLaH10-PRL2019, ExpLaH10-Nature2019}. The lower $T_c$ in CeH$_9$ is associated with relatively lower EPC constant compared to the case of LaH$_{10}$~\cite{LaH10-liangliang,chongze1,chongze2}. It is also noted that the H-derived DOS of CeH$_9$ at $E_F$ is smaller than that of LaH$_{10}$ [see fig. 4(a)].

\begin{figure}[h!t]
\centering{ \includegraphics[width=8.5cm]{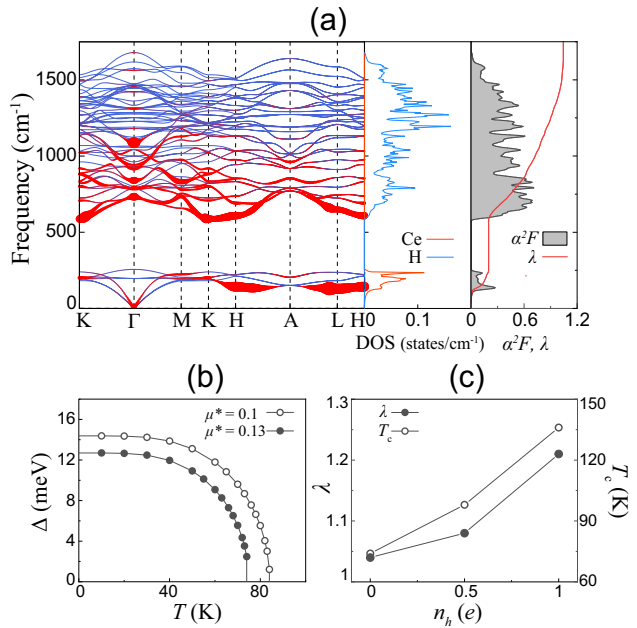} }
\caption{(a) Calculated phonon spectrum, phonon DOS projected onto Ce and H atoms, Eliashberg function ${\alpha}^{2}F({\omega})$, and integrated EPC constant ${\lambda}({\omega})$ of CeH$_9$, (b) superconducting energy gap ${\Delta}$ as a function of temperature with ${\mu}^*$ = 0.1 and 0.13, and (c) ${\lambda}$ and $T_c$ as a function of $n_h$.}
\end{figure}

Since a vHs in the electronic DOS of CeH$_9$ is located below $E_F$ [see Fig. 1(b)], hole doping is expected to induce a shift of $E_F$ toward the vHs. The calculated band structure at a hole doping of $n_h$ = 1.0$e$ per Ce atom shows that $E_F$ approaches the vHs, thereby giving rise to an increase of DOS around $E_F$ (see Fig. S4 in the Supplemental Material~\cite{SM}). In order to examine how the hole doping influences SC, we use the isotropic Migdal-Eliashberg formalism~\cite{Migdal,Eliash,ME-review} to estimate the variations of ${\lambda}$ and $T_c$ as a function of $n_h$. As shown in Fig. 3(c), ${\lambda}$ is enhanced from 1.04 (without hole doping) to 1.08 and 1.21 at $n_h$ = 0.5 and 1.0$e$, respectively, which in turn increases $T_{\rm c}$ up to 136 K at $n_h$ = 1.0$e$. It is thus likely that the increased DOS around $E_F$ via hole doping increases the EPC channels, resulting in an increase of $T_{\rm c}$. We note that the hole doping with $n_h$ $<$ 1.2$e$ preserves structural stability without imaginary phonon frequencies (see Fig. S5 in the Supplemental Material~\cite{SM}).

Although ordinary hole doping is achieved by the introduction of electron acceptor dopants in the host matrix, we here propose the hole doping of CeH$_9$ using the substitution of Ce$^{3+}$ ions for Ce atoms. The percentage of Ce$^{3+}$ could change hole-doping concentrations to tune the DOS at $E_F$. Note that the hole doping of $n_h$ = 1.0$e$ can be enabled by the substitution of 33\% Ce$^{3+}$. In order to examine the electronic miscibility of Ce$^{3+}$ ions in the CeH$_9$ matrix, we calculate the charge density of CeH$_9$ without hole doping (see Fig. S6 in the Suppelemental Material~\cite{SM}). Interestingly, we find that the total charge inside the Ce muffin-tin sphere with radius 1.40 {\AA} is 9.55$e$ with including the 5$s^{2}$5$p^{6}$ semicore electrons, close to that (9.63$e$) obtained at $n_h$ = 1.0$e$. This nearly invariance of Ce charges between the two systems implies that both Ce and Ce$^{3+}$ could behave similarly as cations without hole localization. Based on our results for the charge distribution and structural stability of hole doping $n_h$ $<$ 1.2$e$, Ce$^{3+}$ ions are most likely to be electronically miscible with Ce atoms in the CeH$_9$ matrix. We note that in the present calculations, the hole charges are compensated by uniform background charge to maintain charge neutrality, as implemented in the VASP code~\cite{vasp1,vasp2}. This simulation of hole doping is believed to properly describe the incorporated Ce$^{3+}$ ions in the CeH$_9$ matrix, because both Ce and Ce$^{3+}$ with similar cation characters can be equally screened by their surrounding anionic H cages.

Finally, we also explore the hole-doping effect on SC in a recently observed~\cite{ExpLaH10-PRL2019, ExpLaH10-Nature2019} rare earth hydride LaH$_{10}$. As shown in Fig. 4(a), this hydride has a vHs near $E_F$ with a strong hybridization of La 4$f$ and H 1$s$ orbitals~\cite{LaH10-liangliang}, similar to the characteristic of vHs in CeH$_9$ [see Fig. 1(b)]. The calculated ${\lambda}$ and $T_{\rm c}$ values of LaH$_{10}$ are displayed as a function of $n_h$ in Fig. 4(b). Since the DOS around $E_F$ increases with hole doping [see Fig. 4(a)], ${\lambda}$ increases monotonously with increasing $n_h$. Consequently, hole doped LaH$_{10}$ raises $T_c$ from 233 K (without hole doping) to 245 K at $n_h$ = 0.3$e$. Here, the hole doping-induced increase of $T_c$ is 12 K, much smaller than the corresponding ${\Delta}T_c$ ${\approx}$ 62 K in CeH$_9$ [see Fig. 3(c)]. The relatively smaller value of ${\Delta}T_c$ in LaH$_{10}$ is partly associated with the weak variation of DOS around $E_F$ via hole doping [see Fig. 4(a)].

\begin{figure}[htb]
\centering{ \includegraphics[width=8.5cm]{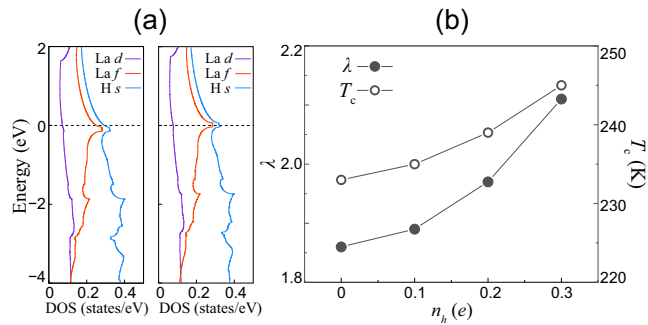} }
\caption{(a) Calculated PDOS of LaH$_{10}$, obtained without hole doping (left) and at $n_h$ = 0.1$e$ (right). The unit of DOS is states/eV per unit cell that contains one La atom. In (b), the calculated ${\lambda}$ and $T_c$ values are displayed as a function of $n_h$.}
\end{figure}

In summary, based on first-principles calculations, we proposed that hole doping significantly enhances $T_c$ in a recently synthesized~\cite{ExpCeH9-Nat.Commun2019T.Cui,ExpCeH9-Nat.Commun2019-J.F.Lin} hydride CeH$_9$. It was revealed that hole doping induces the shift of $E_F$ toward a vHs, thereby leading to the enhancement of EPC through an increased DOS around $E_F$. Interestingly, the vHs was found to be created by the band touching of two crystalline symmetry-protected DNLs whose electronic states are mostly composed of hybridized Ce 4$f$ and H 1$s$ orbitals. Therefore, the symmetry, band topology, and hole doping are cooperated to increase $T_c$ of compressed CeH$_9$. The proposed hole-doping effect on SC is rather generic and hence, it can also be applicable to another experimentally observed~\cite{ExpLaH10-PRL2019, ExpLaH10-Nature2019} high-$T_c$ rare earth hydride LaH$_{10}$. We anticipate that future experimental work will be stimulated to adopt hole doping for rasing $T_c$ in high-pressure superconducting hydrides.

\vspace{0.4cm}

\noindent {\bf Acknowledgements.}
This work was supported by the National Research Foundation of Korea (NRF) grant funded by the Korean Government (Grants No. 2019R1A2C1002975, No. 2016K1A4A3914691, and No. 2015M3D1A1070609). The calculations were performed by the KISTI Supercomputing Center through the Strategic Support Program (Program No. KSC-2020-CRE-0163) for the supercomputing application research.  \\

C.W. and S.L. contributed equally to this work. \\


\noindent $^{*}$ Corresponding author: chojh@hanyang.ac.kr

\end{document}